\documentclass[prl,showpacs,floatfix,amsmath,amssymb,superscriptaddress]{revtex4-1}
\usepackage{amsfonts}
\usepackage{mathrsfs}
\usepackage{tipa}
\usepackage{amssymb}
\usepackage{txfonts}
\usepackage{graphicx}
\usepackage{dcolumn}
\usepackage{epstopdf}
\usepackage[colorlinks,linkcolor=blue,urlcolor=blue,citecolor=blue]{hyperref}
\usepackage{multirow}
\usepackage{subfigure}
\usepackage{url}

\begin{document}
\newcommand*{\cm}{cm$^{-1}$\,}
\newcommand*{\Tc}{T$_c$\,}

\title{Photoinduced hidden CDW state and relaxation dynamics of 1T-TaS$_{2}$ probed by time-resolved terahertz spectroscopy}
\author{Z. X. Wang}
\affiliation{International Center for Quantum Materials, School of Physics, Peking University, Beijing 100871, China}

\author{Q. M. Liu}
\affiliation{International Center for Quantum Materials, School of Physics, Peking University, Beijing 100871, China}

\author{L. Y. Shi}
\affiliation{International Center for Quantum Materials, School of Physics, Peking University, Beijing 100871, China}

\author{S. J. Zhang}
\affiliation{International Center for Quantum Materials, School of Physics, Peking University, Beijing 100871, China}

\author{T. Lin}
\affiliation{International Center for Quantum Materials, School of Physics, Peking University, Beijing 100871, China}

\author{T. Dong}
\affiliation{International Center for Quantum Materials, School of Physics, Peking University, Beijing 100871, China}

\author{D. Wu}
\affiliation{International Center for Quantum Materials, School of Physics, Peking University, Beijing 100871, China}

\author{N. L. Wang}
\email{nlwang@pku.edu.cn}
\affiliation{International Center for Quantum Materials, School of Physics, Peking University, Beijing 100871, China}
\affiliation{Collaborative Innovation Center of Quantum Matter, Beijing, China}

\begin{abstract}
The dynamical properties of single crystal 1T-TaS$_{2}$ are investigated both in commensurate charge density wave state (CCDW state) and hidden charge density wave state (HCDW state). We develop a useful criterion in time-domain transmission terahertz measurement to judge whether the compound is driven into a metastable state or still in its virgin state. An increase of terahertz conductivity by two orders of magnitude from CCDW state to HCDW state is obtained by taking account of the penetration depth mismatch, which is in agreement with reported \emph{dc} transport measurement. Upon weak pumping, only transient processes with rapid decay dynamics are triggered in both CCDW and HCDW states. We compare the conductivity increases in terahertz frequency range between transient and HCDW states and suggest that fluctuated metallic domain walls may develop in the transient states.

\end{abstract}


\maketitle

Ultrafast optical spectroscopy is a powerful tool to investigate the non-equilibrium physics in complex electronic materials. Ultrafast photoexcitations can create two significantly different states. One is the transient nonequilibrium state, which is usually triggered by weak excitations and characterized by a distribution of excited quasiparticles and a rapid relaxation process back to the equilibrium state \cite{Koshihara1990,Yu1991,Cavalleri2001,Tomimoto2003,Okamoto2004,Takubo2005}. The other is the metastable state or even stable phase, which is induced by strong pump pulses. Creation of metastable/stable state opens a new avenue for ultrafast optical manipulation and control of electronic properties of quantum materials \cite{Chakravarty1983,Miyano1997,Kiryukhin1997,Fiebig1925,Takubo2005,Stojchevska2014,Giannetti2016}. More interestingly, some of the metastable/stable states induced by ultrashort laser pulses can even not be accessed in equilibrium state via a change of temperature or pressure. Optical manipulations and ultrafast switching  represent a rapid developing frontier in condensed matter physics.

1T-TaS$_{2}$, a well-known protype charge density wave (CDW) compound, has both transient state and metastable/stable state upon excitations by ultrashort laser pulses. 1T-TaS$_{2}$ has a two dimensional structure consisting of stacked layers of triangular lattice of Ta sandwiched between layers of S atoms with octahedral coordination. Upon cooling, the compound experiences several phase transitions. It changes from a metal to an incommensurate CDW state at 550 K, then transforms into a nearly commensurate CDW at 350 K, where the lattice forms regular array of star-shaped clusters of Ta atoms, named as ``star of David", separated by domain walls \cite{doi:10.1080/00018737500101391}. The star of David cluster can be described by the displacement of 12 Ta atoms towards ``star center'' Ta atoms. Below 180 K, the domain walls vanish and a fully commensurate hexagonal star superlattice forms. The commensurate CDW (CCDW) state is believed to be a Mott insulator with an energy gap of about 100 meV as one electron is localized at the Ta star cluster \cite{doi:10.1080/13642817908245359,Clerc_2007}. When the 1T-TaS$_{2}$ is exposed to a train of strong ultrafast femtosecond laser pulses, the sample can be driven into a so-called ``hidden CDW state (HCDW)" which is actually a stable phase and characterized by a sudden drop of about two to three orders of magnitude in \emph{dc} resistivity at low tempreature \cite{Stojchevska2014}. Time resolved transmission electron microscopy revealed formation of new domain structure in the HCDW state with modulation wave vector different from thermal induced phases \cite{Suneaas9660}.

Up to now, available time-resolved spectroscopy experiments on 1T-TaS$_{2}$ mainly focused on the relaxation dynamics of pump induced nonequilibrium state and the excitation of coherent CDW amplitude-mode oscillations \cite{Demsar2002,Toda2004,Perfetti_2008,Hellmann2010, Hellmann2010, Petersen2011, Kusar2011, Ishizaka2011, EICHBERGER20139}. The photoinduced HCDW state was investigated mainly by \emph{dc} transport and structural characterizations. A previous time-resolved terahertz (THz) spectroscopy combined with pump-probe reflection measurement at several fixed frequencies provided information about collapse and rapid recovery of the Mott gap within a few picoseconds \cite{Dean2011}. However, the conductivity in the THz frequency range increases by only a factor of 2$\sim$3 after photoexcitations, which is very different from the \emph{dc} resistivity measurement with a magnitude change about two to three orders between CCDW state and HCDW state\cite{Stojchevska2014}. The measurement might be limited to the transient state of CCDW state owing to the weak pump fluence. The dynamical properties in the HCDW state remain unexplored up to present. Here, we present the low frequency dynamics in both CCDW and HCDW states by performing time domain THz spectroscopy measurement in transmission configuration. By shining intense nearinfrared laser pulses on very thin 1T-TaS$_{2}$ single crystal sample, we are able to drive the sample from CCDW state to HCDW state. An increase of THz conductivity by two orders is obtained after taking account of the penetration depth mismatch between near infrared pump and THz probe beams. We also present the relaxation dynamics of the transient states in both CCDW and HCDW states.

High-quality single crystals of 1T-TaS$_{2}$ are synthesized by chemical vapor transport method in a sealed quartz tube followed by a quenching process to retain the 1T phase (more details in \cite{Wu2018}). The crystals were further cleaved enabling to obtain very thin samples with thickness about 5 $\mu m$ determined by a step profiler. The equilibrium and photoexcitation-induced spectra are measured in a home-built time-resolved terahertz transmission spectroscopy system at 6 K. The electric field of a THz pulse that passes through a sample or the same size aperture (1 mm of diameter) as reference is recorded as a function of time delay. Fourier transformation of the recorded time traces provides the frequency dependent complex transmission spectra which contains both magnitude and phase information. A 800 nm pulse with tunable power intensity also from the amplified laser is used as pump beam to drive a phase transition from CCDW state to HCDW state or to induce transient nonequilibrium state (more details are provided in supplementary \cite{Supplemental}).

\begin{figure}[htbp]
	\centering
	\includegraphics[width=8cm]{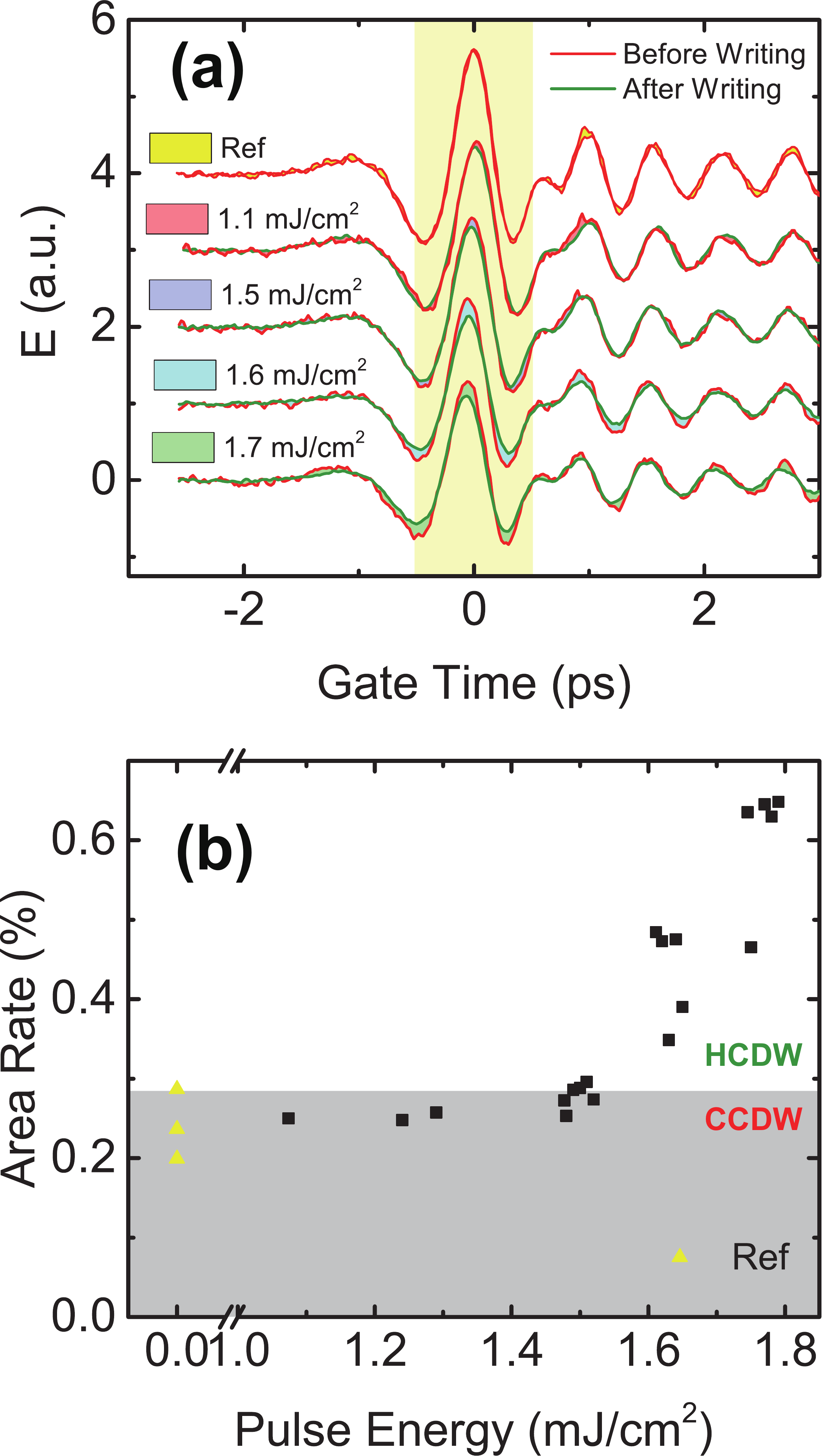}\\
	\caption{
(a) THz transmission electric fields of 1T-TaS$_{2}$ crystal measured at 6 K before and after excitation pulses with different fluences. (b) The sum of the spectral difference between two THz spectra before and after excitations normalized by the area under the electric field curve in the yellow region shown in (a) at different fluences. A threshold of about 1.5 mJ/cm$^2$ can be identified.}\label{Fig:1}
\end{figure}

To distinguish between the two remarkable different states by laser excitations, it is essential to introduce a useful criterion to judge whether the 1T-TaS$_{2}$ has been driven into HCDW state by ultrashort excitation pulses or still in its virgin CCDW state. The strong excitation pulse leading to the phase switching is defined as the ``write" pulse, while the weak excitation pulse that induces only transient state is still simply called ``pump" pulse. Then, we can identify different states directly by optical method without fabricating sophisticated devices and performing \emph{dc} transport measurement. We first measure the THz transmission electric field in time domain of the 1T-TaS$_{2}$ sample in CCDW state at 6 K without any excitation pulses, that is, the sample is in the dark (the red curve in Fig.\ref{Fig:1} (a)). After illuminating on the sample for a second, the excitation pulses are blocked and the transmission electric field is measured again (the green curve in Fig.\ref{Fig:1} (a)). If the excitation pulses are strong enough above a threshold, 1T-TaS$_{2}$ is driven from CCDW state into HCDW state, leading to a reduction of the transmitted THz electric field due to suddenly enhanced conductivity. Then, the difference between the red curve and the green curve becomes eminent, as displayed most clearly in the two bottom cases in Fig.\ref{Fig:1} (a). By contrast, when the intensity is below the threshold, the difference between the two curves is contributed mainly from random noise, then the sample remains in its virgin CCDW state. In order to distinguish the phase transition from trivial random noise, we plot the sum of the difference between the measured curves normalized by the area under the electric field curve as a function of laser excitation fluence. As the difference is most clearly observed near the main peak of THz signal, we summarize the spectral difference only this range from -0.5 ps to 0.5 ps (the yellow region in Fig.\ref{Fig:1} (a)). The result is shown in Fig.\ref{Fig:1} (b). The normalized difference between two curves without writing pulse can never be more than 0.29\%. The difference can go beyond this value when the pulse energy reaches 1.5 mJ/cm$^{2}$. Consequently, 1T-TaS$_2$ is driven to a stable HCDW state only when the pulse intensity is higher than this threshold. In the subsequent part of this work, 0.8 mJ/cm$^{2}$ is chosen for the pump pulse, while 1.8 mJ/cm$^{2}$ serves as the writing pulse.

\begin{figure}[htbp]
	\centering
	\includegraphics[width=10cm]{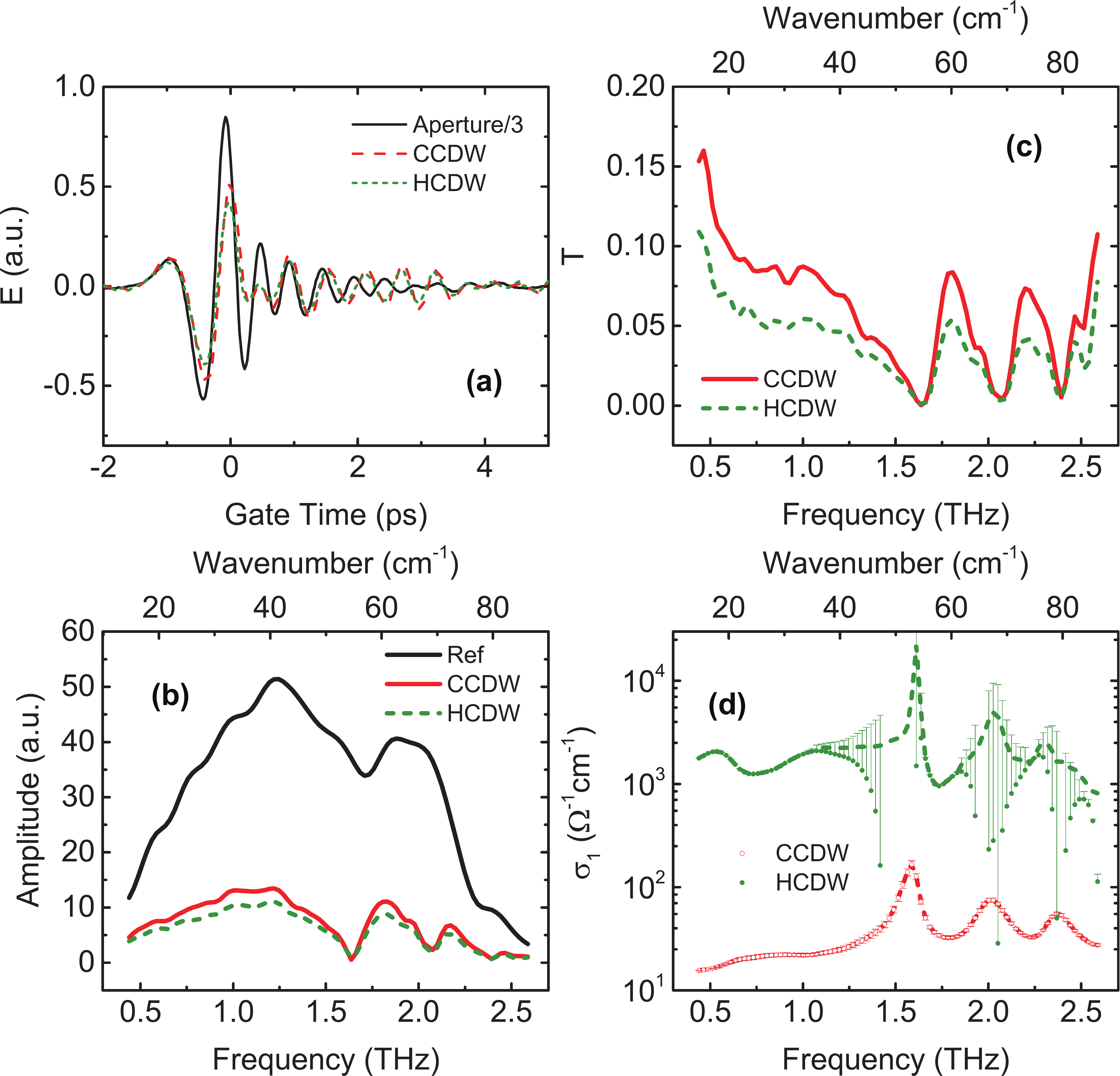}\\
	\caption{(a) The transmitted THz electric field of CCDW state (red dash curve) and HCDW state (green dot curve) in 1T-TaS$_{2}$ at 6 K. Black curve represents THz electric field passing through an empty aperture. The HCDW is achieved by strong writing pulses (1.8 mJ/cm$^{2}$). (b) The spectra in frequency domain via Fourier transformation of time-domain THz curves in (a). (c) The transmission spectra ${T^{\rm{CCDW}}=\left| {{{\tilde E}^{\rm{CCDW}}}} \right|^{2}/\left| {{{\tilde E}^{\rm{ref}}}} \right|^{2}}$ and ${T^{\rm{HCDW}}=\left| {{{\tilde E}^{\rm{HCDW}}}} \right|^{2}/\left| {{{\tilde E}^{\rm{ref}}}} \right|^{2}}$. (d) The real part of conductivity of equilibrium CCDW and HCDW states, respectively. The dash lines are guided to eyes.}\label{Fig:2}
\end{figure}

Figure \ref{Fig:2} (a) and (b) present the time traces of static THz electric field $E\left( t \right)$ and its Fourier transform $\left| {\tilde E\left( \omega  \right)} \right|$ passing through the empty aperture as a reference (black curves) and the thin sample (red curves) at 6 K, respectively. The THz signal passing through the sample is significantly reduced, indicating that the sample thickness is close to the penetration depth in THz regime. In fact, the THz signal is almost absent at room temperature due to its metallic property. With a strong writing pulse, the sample can be induced from CCDW state into HCDW state at 6 K. The corresponding electric field and Fourier transformed spectra in HCDW state are plotted as the green dash curves in Fig.\ref{Fig:2} (a) and (b), respectively. The signal intensities are further reduced. Three phonon modes (1.6 THz, 2.1 THz , 2.4 THz) are observed both in CCDW state and HCDW state. They contribute to the oscillations in time domain measurement in Fig.\ref{Fig:2} (a), which can last longer than 15 ps and decay very slowly. Those phonon modes were reported in far-infrared measurement by Fourier transform infrared spectrometer \cite{Gasparov2002} and previous THz reflection measurement \cite{Dean2011}. The transmission spectra ${T^{\rm{CCDW}}\left( \omega  \right)=\left| {{{\tilde E}^{\rm{CCDW}}\left( \omega  \right)}} \right|^{2}/\left| {{{\tilde E}^{\rm{ref}}\left( \omega  \right)}} \right|^{2}}$ and ${T^{\rm{HCDW}}\left( \omega  \right)=\left| {{{\tilde E}^{\rm{HCDW}}\left( \omega  \right)}} \right|^{2}/\left| {{{\tilde E}^{\rm{ref}}\left( \omega  \right)}} \right|^{2}}$ are shown in Fig. \ref{Fig:2} (c). The complex optical electric conductivity of CCDW state can be obtained from $\tilde t(\omega ) = {\tilde E^{{\rm{Sample}}}}(\omega )/{\tilde E^{{\mathop{\rm Re}\nolimits} {\rm{f}}}}(\omega )$, by the relation (see supplementary \cite{Supplemental}):
\begin{equation}
\tilde t\left( \omega  \right) = \frac{{4\tilde n\left( \omega  \right) \cdot \exp \left[ {{\rm{i}}{k_0}d\left( {\tilde n\left( \omega  \right) - 1} \right)} \right]}}{{{{\left( {1 + \tilde n\left( \omega  \right)} \right)}^2} - {{\left( {\tilde n\left( \omega  \right) - 1} \right)}^2}\exp \left( {2{\rm{i}}{k_0}d\tilde n\left( \omega  \right)} \right)}}
\label{con:equilibriumstate}
 \end{equation}
where d is the sample thickness (5 $\mu$m determined by the step profiler). To obtain the conductivity of HCDW state, the mismatch of penetration depths between write and probe pulses should be taken into account (approximately 50 nm and 8 $\mu $m for 800 nm pump beam and THz probe range, respectively). Since the penetration depth of the write pulse is much shorter than the sample thickness, only a small part can be induced into HCDW state while other part is still in CCDW state. This mismatch effect is just like a thin film in HCDW state coated on a substrate in CCDW state (see supplementary for calculation process in details \cite{Supplemental}). We note that the measured transmission data near the three phonon modes are closed to zero, a small level of noise would lead to a large uncertainty in conductivity in the phonon mode frequencies. Nevertheless, those are not our focus in this work. Our calculated conductivity in the THz regime shows clearly a dramatic change of about two orders in magnitude, which is thus in agreement with the change observed in \emph{dc} conductivity \cite{Stojchevska2014}.

\begin{figure}[htbp]
	\centering
	\includegraphics[width=10cm]{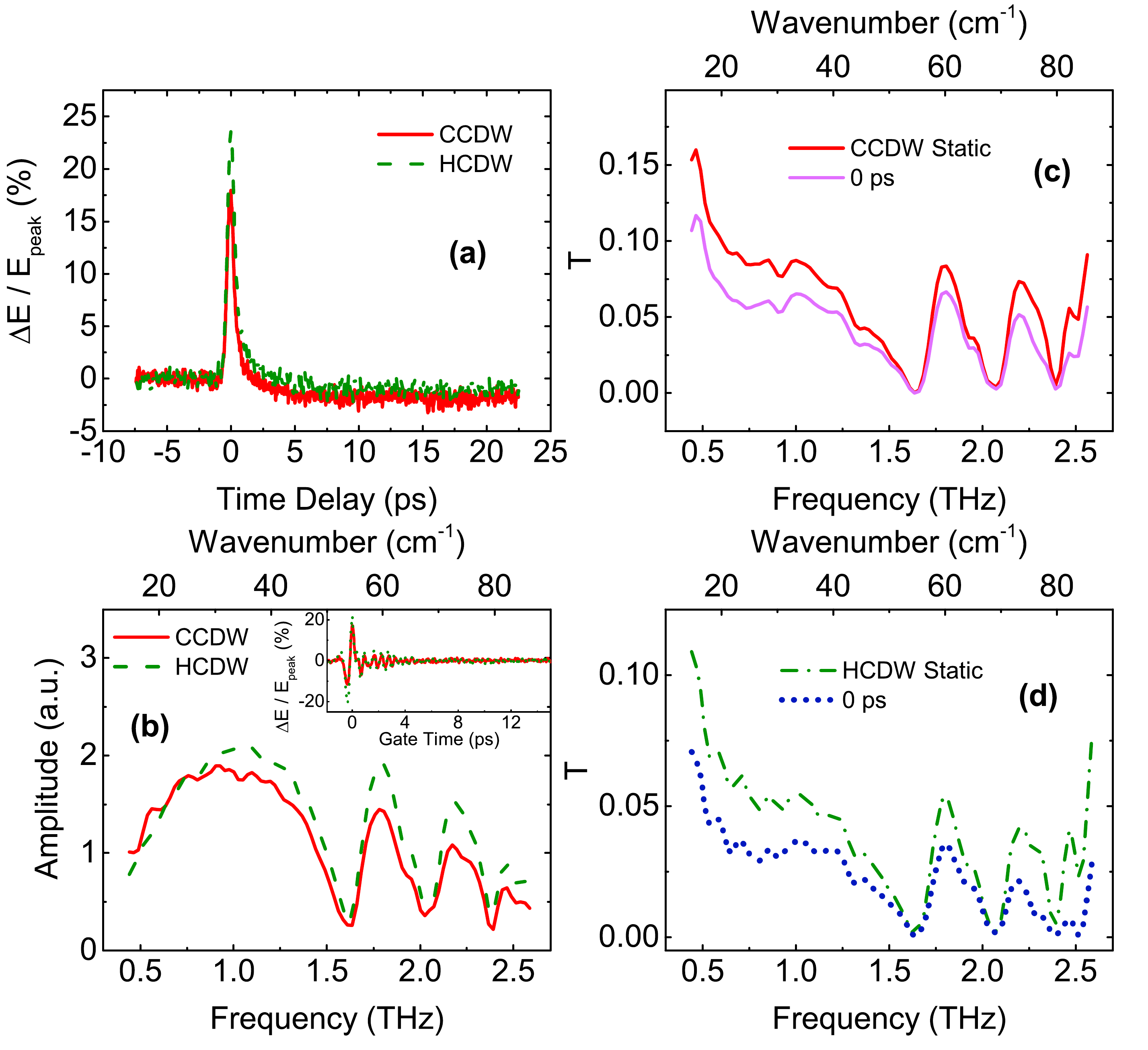}\\
	\caption{(a) The relative change of the transmitted THz electric field at different decay time after excited by 800 nm pulses at fluence of 0.8 mJ/cm$^{2}$. The red and green curves show the decay of CCDW and HCDW states, respectively. (b) Inset: $\Delta E\left( {t,\tau =0} \right)/{E_{{\rm{peak}}}}$. Main panel: the Fourier transformed spectra from $\Delta E ({t,\tau =0})$. (c) and (d) The transmission spectra before and after excitation by 0.8 mJ/cm$^{2}$ pulses from CCDW state and HCDW, respectively. }\label{Fig:3}
\end{figure}

In an earlier time-resolved time domain reflection terahertz experiment on 1T-TaS$_2$, a small pump-induced change in THz conductivity is observed. The conductivity increases by only a factor of 2$\sim$3 at the maximum pump-induced signal and relaxes back to the equilibrium values very fast \cite{Dean2011}. The small change and the rapid relaxation process are not compatible to the two to three orders of magnitude change of conductivity from CCDW to HCDW state. It is also much smaller than the effect obtained in this study. It is possible that the pump excitations did not drive a phase transition from CCDW to HCDW due to the small pump fluence about 0.55-mJ/cm$^2$ used in the reported work \cite{Dean2011}, which is much lower than the switching threshold in our experiment. Nonetheless, even if the pump fluence was higher than the threshold, the measurement method could still not access to the static response of the HCDW state. This is because the pump laser beam is always shining on the sample in such measurement. In this circumstance, what is measured is the pump induced change in the HCDW state.

To further study the photo-induced nonequilibrium transient states and their relaxation dynamics in both CCDW and HCDW phases, we performed time-resolved THz measurement on the sample in those two states with a pump fluence less than the switching threshold. Figure \ref{Fig:3} (a) displays the decay process of the pump-induced changes of the THz electric field at peak position $\Delta E(t)/E_{\rm{peak}}$ for both CCDW state and HCDW state after excitation at 800 nm at a fluence of 0.8-mJ/cm$^{2}$. Even though the equilibrium states for CCDW state and HCDW state are quite different, the relaxation behaviors are similar: after a sudden increase near zero time delay, they relax exponentially and back to initial state roughly within 3 ps. The process is totally different from the switching effect to a stable new phase driven by strong write pulses \cite{Stojchevska2014,Vaskivskyie1500168}. The inset and the main panel in Fig.\ref{Fig:3} (b) are the time dependent relative change of the THz electric field $\Delta E(t, \tau = 0 \rm{ps})/E_{\rm{peak}}$ and the Fourier transformation of $\Delta E(\omega, \tau = 0 \rm{ps})$ at the gate delay time $\tau =0$ ps (where the pump-induced change of electric field reaches the maximum value as displayed in Fig.\ref{Fig:3} (a)), respectively. The oscillations in the inset of Fig.\ref{Fig:3} (b) is due to phonon modes. For both CCDW state and HCDW state, $\Delta E(\omega, \tau = 0 \rm{ps})$ appear similar to the spectrum in the equilibrium state $ E(\omega)$ as shown in Fig.\ref{Fig:2} (b), except for the much smaller values.  The transmission spectra ${T( \omega, \tau = 0 \rm{ps})=\left| {{{\tilde E}\left( \omega  \right)}} +\Delta E(\omega, \tau = 0 \rm{ps}) \right|^{2}/\left| {{{\tilde E}^{\rm{ref}}\left( \omega  \right)}} \right|^{2}}$ for CCDW state and HCDW state are shown in Fig.\ref{Fig:3} (c) and (d), respectively.

   \begin{figure}[htbp]
 	\centering
 	\includegraphics[width=10cm]{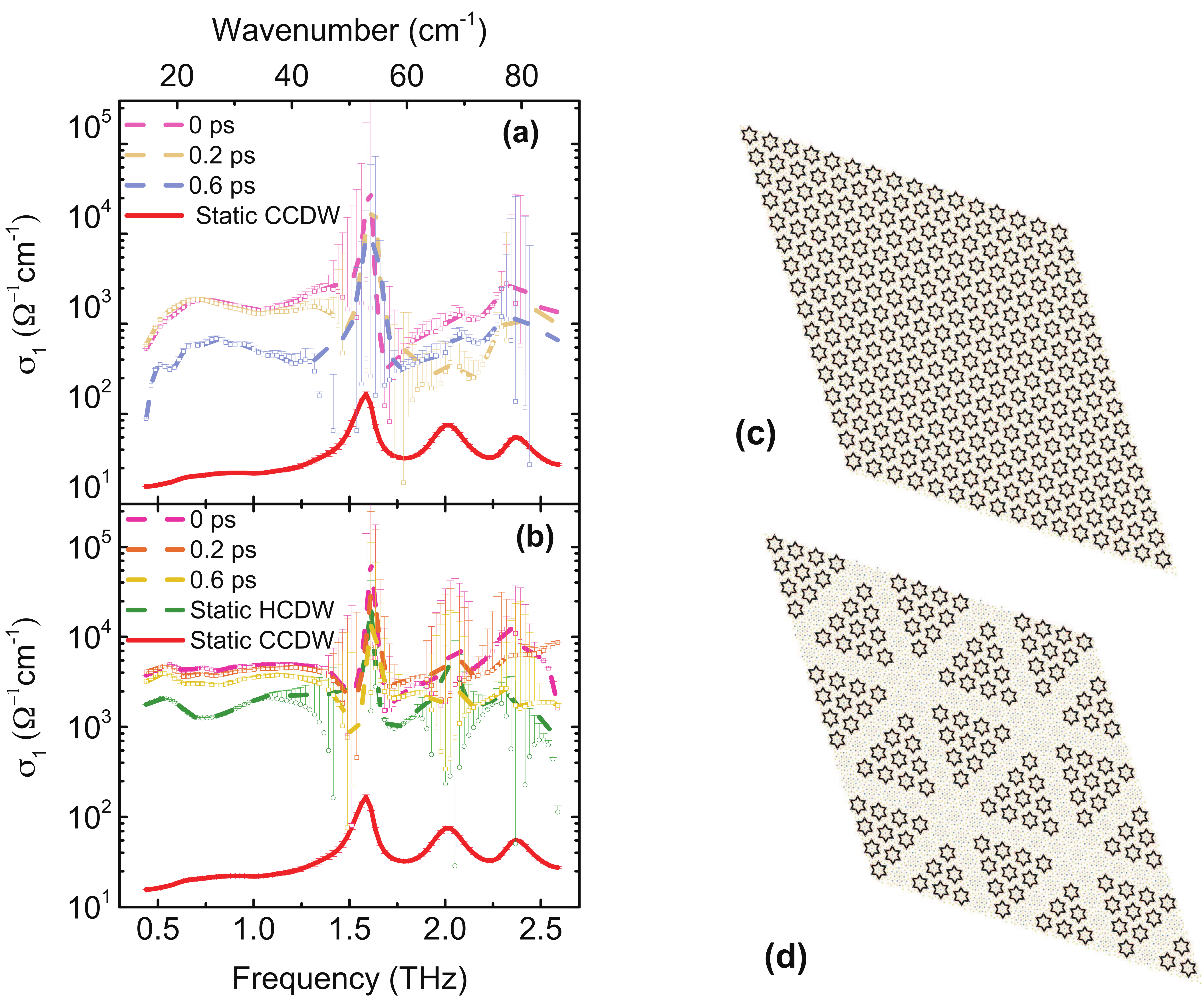}\\
 	\caption{ (a) and (b) Real part of conductivity as a function of time delay after excitations by a pump fluence of 0.8 mJ/cm$^{2}$ for CCDW and HCDW states, respectively. The dash lines are guided to eyes. (c) David-star pattern in the CCDW state. (d) Formation of metallic domain walls between David-star clusters in the HCDW sate.}\label{Fig:4}
 \end{figure}

The mismatch of the penetration depths between pump and probe beams has to be considered again. The penetration depth of the pump is the same as that of the write pulse (50 nm). We obtain the transient optical conductivities for both CCDW state and HCDW state at different delay times, as shown in  Fig.\ref{Fig:4} (a) and (b), respectively. The weak near-infrared pump induces a transient state. The rapid relaxation process back to the equilibrium state is similar to the earlier time-resolved THz measurement \cite{Dean2011}. However, the increase of conductivity at the time delay $\tau =0$ ps excited from the CCDW state is much higher than the photoinduced conductivity reported in earlier work \cite{Dean2011}. The values are close to that in the HCDW state. This is reasonable as we shall explain below that the transient state represents fluctuated situation with metallic domain walls. The sudden increase of conductivity suggests a melting of the Mott gap in the CCDW state. Similar transient behaviors are also observed in the HCDW state, but the conductivity increase relative to the static HCDW state is smaller in comparison to the transient excitations from the CCDW state. The smaller increase of conductivity in the HCDW state could be easily understood since the HCDW state has much higher conductivity than CCDW state. Due to the limited frequency range, the conductivity values are roughly frequency independent except for the phonon modes.

We now discuss the physical pictures achieved in the different states probed by our time resolved THz measurement. In the CCDW state, the conductivity in the THz range is very low. The compound is in a Mott-insulator state with the lattice forming a regular array of star of David without domain walls, as displayed in Fig.\ref{Fig:4} (c). In the HCDW state, long range ordered domain walls are created by the laser excitations \cite{Stojchevska2014,Suneaas9660}. The metallic channels are formed in the domain walls, leading to the sudden increase of conductivity by two orders in magnitude, as shown in Fig.\ref{Fig:4} (d). While in the transient state excited in the CCDW state, the increase of conductivity is close to the values of HCDW state, which suggests that the transient state is similar to the HCDW state, likely with development of fluctuated metallic domain walls in the David-star pattern. Those fluctuated domain walls relax back to the equilibrium CCDW state quickly, roughly within 3 ps. The transient state excited from the HCDW state may also have fluctuated domain walls. As the transient conductivity values are higher than the HCDW state, the fluctuated domain walls may have different modulation wave vectors.

To summarize, the transmission THz study on thin single crystals demonstrates a useful way to distinguish the dynamical properties between photoinduced metastable/stable state and transient nonequilibrium state. The strong laser pulses drive a phase transition of 1T-TaS$_{2}$ compound from CCDW state to HCDW state, being characterized by a reduction of the transmitted THz electric field. The HCDW is a stable phase at low temperature at 6 K. This allows to measure and determine the optical conductivities in the THz frequency range by the static time-domain measurement. An increase of THz conductivity by two orders, after taking account of the penetration depth mismatch, is in good agreement with the \emph{dc} transport measurement on fabricated small device induced by strong photoexcitations. On the other hand, the weak laser pulses induce transient processes with rapid decay dynamics in both CCDW and HCDW states. A similar conductivity increase in THz frequency range at the maximum pump-induced change enables us to propose that some fluctuated metallic domain walls develop in the pattern of David-star in those transient states. Finally, we would like to remark that the measurement technique and analysis procedure developed in this study should be also applicable to other electronic systems with photoinduced phase transitions.

\begin{center}
\small{\textbf{ACKNOWLEDGMENTS}}
\end{center}
This work was supported by National Natural Science Foundation of China (No. 11888101), the National Key Research and Development Program of China (No. 2017YFA0302904, 2016YFA0300902).

\bibliographystyle{apsrev4-1}
\bibliography{TaS2}

\clearpage
\begin{appendix}
\begin{center}
{\bf  \Large
Supplementary Information:

Photoinduced hidden CDW state and relaxation dynamic of 1T-TaS$_{2}$ probed by time-resolved terahertz spectroscopy
}  %
\vspace{\baselineskip} %

\small Z. X. Wang$^{1}$, Q. M. Liu$^{1}$, L. Y. Shi$^{1}$,  S. J. Zhang$^{1}$,   T. Lin$^{1}$,  T. Dong$^{1}$,  D. Wu$^{1}$ and N. L. Wang$^{1, 2}$
 \\

{\it
$^{1}$ International Center for Quantum Materials, School of Physics, Peking University, Beijing 100871, China\\
$^{2}$ Collaborative Innovation Center of Quantum Matter, Beijing, China\\
}

\end{center} %

\vspace{1\baselineskip} %

\section*{\large Appendix A: Optical pump-THz probe spectroscopy}
FigS.\ref{Fig:S1} shows the experimental setup of the optical pump-THz probe spectroscopy. We use a regenerative amplified Ti:sapphire laser with the pulse duration of 100 fs, the repetition rate of 1 kHz, and the center wavelength at 800 nm. The terahertz signal, ranging from 0.25 to 2.7 THz, is generated by optical rectification on a (110) ZnTe crystal from 800 nm pulses and detected by 1 mm ZnTe crystal via free space electro-optics sampling.
\\

\setcounter{figure}{0}
\renewcommand\figurename{Fig.S}
\begin{figure}[h]
\begin{center}
\includegraphics[height=7cm]{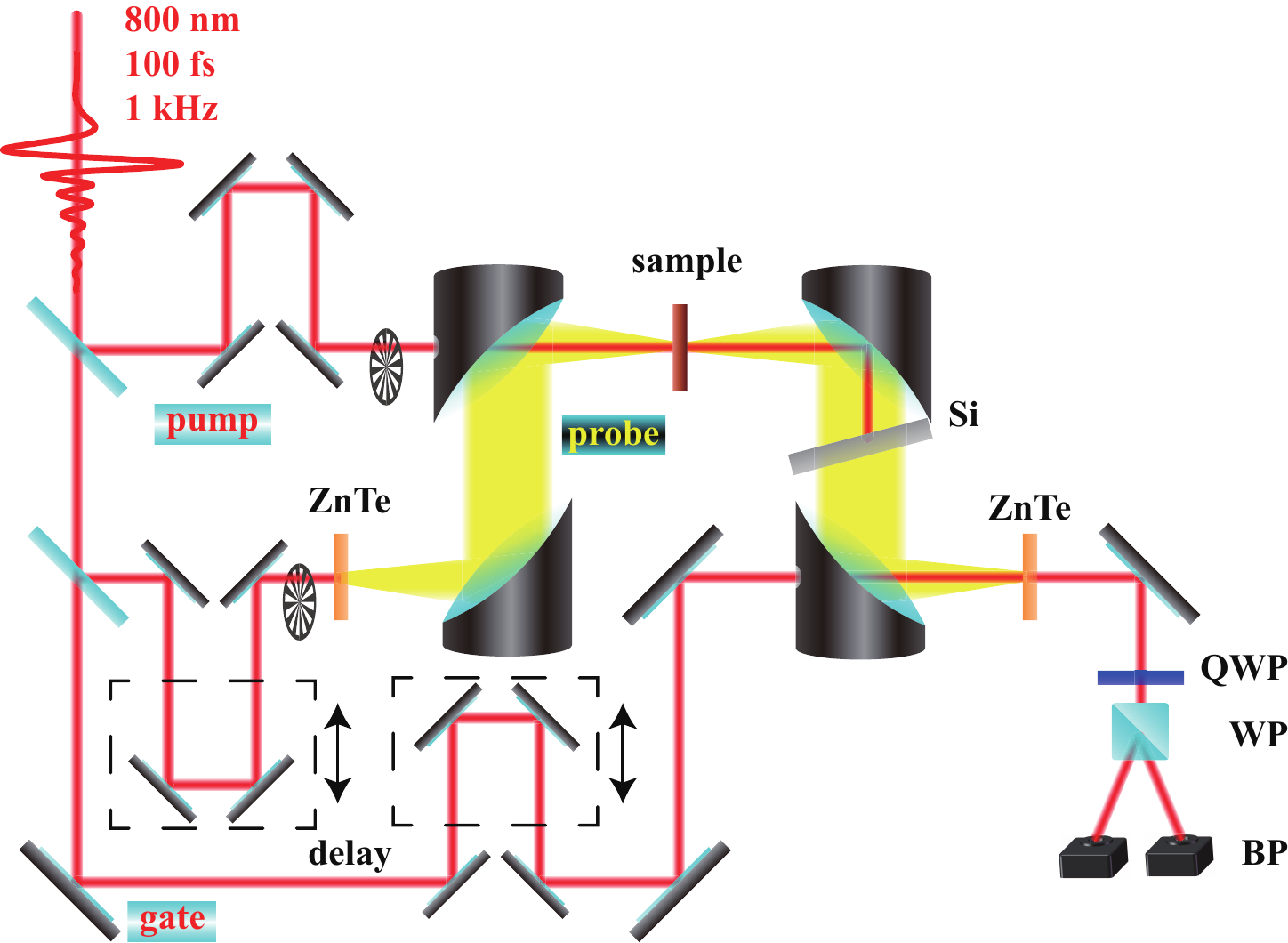}
\vspace{0cm}
\caption{Experimental setup of the optical pump-THz probe spectroscopu. QWP: quarter wave plate, WP: Wollaston prosm, BP: balanced photodiode.}
\label{Fig:S1}
\end{center}
\end{figure}

\section*{\large Appendix B:  Monolayer and multilayer model for optical constants calculation}

\begin{figure}[h]
\begin{center}
\includegraphics[height=5cm]{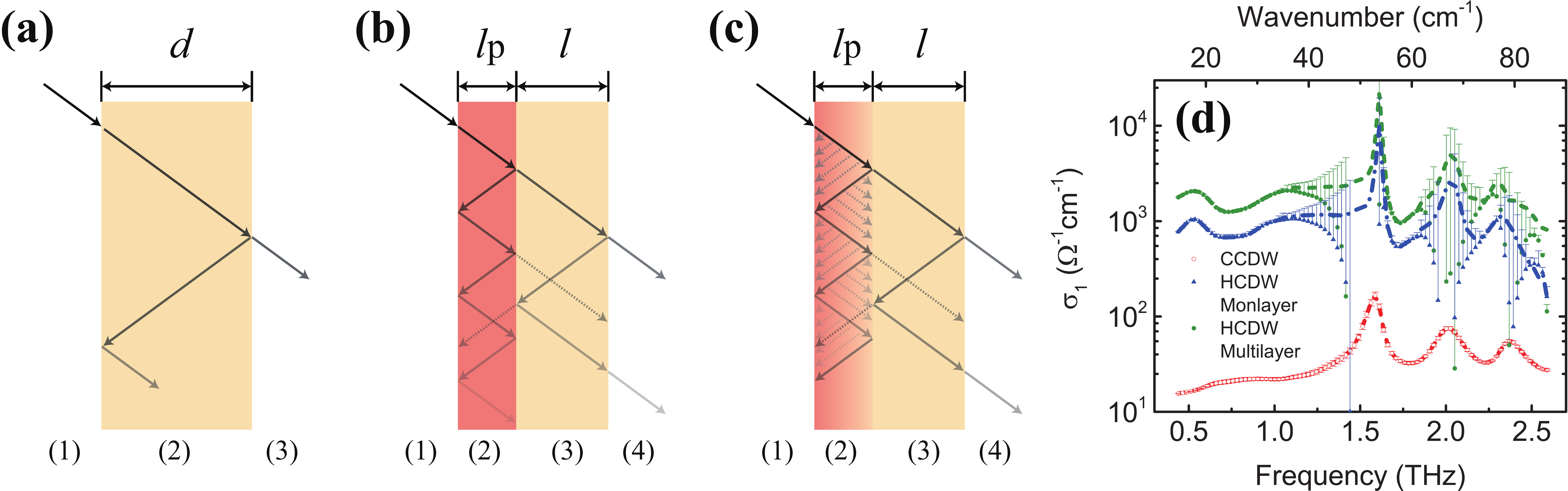}
\vspace{0cm}
\caption{(a)(b)(c) Diagram of thin film, monolayer and multilayer considering interference caused by multireflection, respectively. (d) the real part of conductivity of equilibrium state(open red circles), HCDW state or transient state obtained by monolayer model (blue triangles), and multilayer model(green filled circles).}
\label{Fig:S2}
\end{center}
\end{figure}
\setcounter{equation}{0}
Due to the mismatch of penetration depths between write (or pump) and probe pulses we should take it into account to obtain the conductivity of HCDW state (or transient state). The penetration depth of the write (or pump) pulse is much shorter than the sample thickness, so only a small part changes after pumping while other parts are still in virgin state. Because the 1T-TaS2 sample is very thin and no distinguishable echoes appear in our result, so the the situation about reflection off and transmission through a dielectric slab with thickness d should be considered as shown in Fig.S\ref{Fig:S2} (a). Propagates in a plane parallel film with a thickness of z can be given by the characteristic matrix $\mathbf{M}(z)$ [1]:
 \begin{equation}
\mathbf{M}(z)=\left[ \begin{array}{cc}{\cos(k_{0}\tilde{n}z\cos\theta)} & {-\frac{\mathrm{i}}{\tilde{p}}\sin( k_{0} \tilde{n} z \cos \theta)} \\ {-\mathrm{i} \tilde{p}\sin( k_{0} \tilde{n} z \cos \theta)} & {\cos(k_{0}\tilde{n}z\cos\theta)}\end{array}\right].
\label{equ:matrix}
\end{equation}
where $\theta=0$ for normal incidence, $k_{0} = \omega/c$, and ${\tilde{p}=\sqrt{\frac{\epsilon}{\mu_{0}}}\cos\theta=\tilde{n}}$ for the transverse electric field configuration case. The transmission coefficient of sample obtained by time-domain THz spectroscopy measurements $\tilde{t}$ is
\begin{equation}
\tilde{t}_{\rm{Sample}}=\frac{\tilde{E}_{\rm{Sample}}}{\tilde{E}_{\rm{in}}}=\frac{2\tilde{p}_{1}}{(m_{11}+m_{12}\tilde{p}_{l})\tilde{p}_{1}+(m_{21}+m_{22}\tilde{p}_{l})}
\label{equ:sample}
\end{equation}
and of aperture is
\begin{equation}
\tilde{t}_{\rm{Ref}}=\frac{\tilde{E}_{\rm{Ref}}}{\tilde{E}_{\rm{in}}}=\exp(\mathrm{i}k_{0}d)
\label{equ:ref}
\end{equation}
With $\tilde{p}_{1}=\tilde{p}_{l=3}=1$, we can get the relationship between complex refractive index and index of refraction:
\begin{equation}
\tilde t\left( \omega  \right)=\tilde{t}_{\rm{Sample}}/\tilde{t}_{\rm{Ref}} = \frac{{4\tilde n\left( \omega  \right) \cdot \exp \left[ {{\rm{i}}{k_0}d\left( {\tilde n\left( \omega  \right) - 1} \right)} \right]}}{{{{\left( {1 + \tilde n\left( \omega  \right)} \right)}^2} - {{\left( {\tilde n\left( \omega  \right) - 1} \right)}^2}\exp \left( {2{\rm{i}}{k_0}d\tilde n\left( \omega  \right)} \right)}}
\label{equ:1}
 \end{equation}
we can obtain the conductivity of equilibrium CCDW state, as open red circles in Fig.S\ref{Fig:S2} (d). With a pump or write pulse, a exited film with a thickness of pump penetration depth coated on the sample forms, as in FigS.\ref{Fig:S2} (c) and (d). They represent monolayer model and multilayer model, respectively.

\subsection*{\normalsize Monolayer model}
There are two layers with different refractive $\tilde{n}'$ and $\tilde{n}$. In FigS. \ref{Fig:S2} (b), part (1) and (4) represent vacuum. The thickness of layer (2) and layer (3) are $l_{p}$ (the pump pennetration depth) and $l$. In the case of normal incidence $\cos\theta=1$. The characteristic matrices are
\begin{equation}
\mathbf{M}_{(2)}=\left[ \begin{array}{cc}{\cos(k_{0}\tilde{n}'l_{p})} & {-\frac{\mathrm{i}}{\tilde{n}'}\sin( k_{0} \tilde{n}' l_{p} )} \\ {-\mathrm{i} \tilde{n}'\sin( k_{0} \tilde{n}' l_{p})} & {\cos(k_{0}\tilde{n}'l_{p})}\end{array}\right]
\end{equation}
and
\begin{equation}
\mathbf{M}_{(3)}=\left[ \begin{array}{cc}{\cos(k_{0}\tilde{n}l)} & {-\frac{\mathrm{i}}{\tilde{n}}\sin( k_{0} \tilde{n}l)} \\ {-\mathrm{i} \tilde{n}\sin( k_{0} \tilde{n} l)} & {\cos(k_{0}\tilde{n}l)}\end{array}\right]
\label{equ:M3}
\end{equation}
the toal characteristic matrix is $\mathbf{M}=\mathbf{M}_{(2)}\mathbf{M}_{(3)}$. Hence the elements of it are:
\begin{equation}
\left\{
\begin{array}{l}
m_{11}=\cos(k_{0}\tilde{n}'l_{p})\cos(k_{0}\tilde{n}l)-\frac{\tilde{n}}{\tilde{n}'}\sin(k_{0}\tilde{n}'l_{p})\sin(k_{0}\tilde{n}l)\\[2mm]
m_{12}=-\frac{\mathrm{i}}{\tilde{n}}\cos(k_{0}\tilde{n}'l_{p})\sin(k_{0}\tilde{n}l)-\frac{\mathrm{i}}{\tilde{n}'}\sin(k_{0}\tilde{n}'l_{p})\cos(k_{0}\tilde{n}l)\\[2mm]
m_{21}=-\mathrm{i}\tilde{n}\cos(k_{0}\tilde{n}'l_{p})\sin(k_{0}\tilde{n}l)-\mathrm{i}\tilde{n}'\sin(k_{0}\tilde{n}'l_{p})\cos(k_{0}\tilde{n}l)\\[2mm]
m_{22}=\cos(k_{0}\tilde{n}'l_{p})\cos(k_{0}\tilde{n}l)-\frac{\tilde{n}'}{\tilde{n}}\sin(k_{0}\tilde{n}'l_{p})\sin(k_{0}\tilde{n}l)\\[2mm]
\end{array}
\right.
\end{equation}
With Equ.\ref{equ:sample} and Equ.\ref{equ:ref}, we denote the vacuum as medium 1 so that $\tilde{p}_{1}=\tilde{p}_{l=4}=1$.
The transmission coefficient of the sample $\tilde{t}$ is
 \begin{equation}
\tilde{t} = {\tilde{t}_{{\rm{Sample}}}}/{\tilde{t}_{{\mathop{\rm Re}\nolimits} {\rm{f}}}} = \frac{{2\exp \left( { - {\rm{i}}{k_0}d} \right)}}{\begin{array}{l}
2\cos \left( {{k_0}{\tilde{n}'}{l_{\rm{p}}}} \right)\cos \left( {{k_0}{\tilde{n}}l} \right) - \left( {\frac{{\tilde{n}}}{\tilde{n}'} + \frac{\tilde{n}'}{\tilde{n}}} \right)\sin \left( {{k_0}\tilde{n}'{l_{\rm{p}}}} \right)\sin \left( {{k_0}\tilde{n}l} \right)\\
 - {\rm{i}}\left( {\tilde{n} + \frac{{\rm{1}}}{{\tilde{n}}}} \right)\cos \left( {{k_0}\tilde{n}'{l_{\rm{p}}}} \right)\sin \left( {{k_0}\tilde{n}l} \right) - {\rm{i}}\left( {\tilde{n}' + \frac{1}{{\tilde{n}'}}} \right)\sin \left( {{k_0}\tilde{n}'{l_{\rm{p}}}} \right)\cos \left( {{k_0}\tilde{n}l} \right)
\end{array}}
\end{equation}
the solution of this equation is the refractive index of part (2). The conductivity can be obtained from refractive index, as blue triangles shown in FigS. \ref{Fig:S2} (d).
\subsection*{\normalsize Multilayer model}
As shown in FigS.\ref{Fig:S2} (c), the part (2) is different from monolayer model. In monolayer model the refractive index is a constant $\tilde{n}'$, while in multilayer modelthe refrective index of each pumped layer in part (2) evolving in exponential decay, $\tilde{n}(z) = \tilde{n} + \Delta \tilde{n}\cdot \rm{e}^{-z/l_{p}}$. If each layer is thin enough, the characeristic matrix of each layer can be obtained from Equ.\ref{equ:matrix}:
\begin{equation}
\mathbf{M}_{j}=\left[ \begin{array}{cc}{1} & {-\mathrm{i} k_{0} \delta z_{j}  } \\ {-\mathrm{i} k_{0} \tilde{n}_{j}^{2} \delta z_{j} ,} & {1}\end{array}\right].
\end{equation}
The characteristic matrix of medium in part (2) can be written as a product of all the matrices for every layer,
\begin{equation}
\mathbf{M}_{(2)}=\prod_{j=1}^{N} \mathbf{M}_{j}=\left[ \begin{array}{cc}{1} & {-\mathrm{i} k_{0} B} \\ {-\mathrm{i} k_{0} A} & {1}\end{array}\right].\label{con:1}
\end{equation}
where
\begin{equation}
A=\sum_{j=1}^{N} \tilde{n}^{2}_{j} \delta z_{j}, B=\sum_{j=1}^{N}  \delta z_{j}.
\end{equation}
in the limit $|\delta z_{j}|\to0$, we obtain
\begin{equation}
\mathbf{M}_{(2)}=\left[ \begin{array}{cc}{1} & {-\mathrm{i} k_{0} \mathcal{B}} \\ {-\mathrm{i} k_{0} \mathcal{A}} & {1}\end{array}\right].\label{con:1}
\end{equation}
where
\begin{equation}
\begin{array}{l}
\mathcal{A} =  \int^{l_{p}}_{0} \tilde{n}(z)^{2} \mathrm{d} z=n^2l_p+2nl_p(1-1/\mathrm{e})\Delta n+\frac{1}{2}l_p(1-1/e^2)\Delta n^2,\\[2mm]
\mathcal{B} = \int^{l_{p}}_{0} \mathrm{d} z=l_p
\end{array}
\end{equation}
and $\mathbf{M}_{(3)}$ is the same as the part (3) in monolayer case described by Equ.\ref{equ:M3}.
the elements in the toal characteristic matrix $\mathbf{M}=\mathbf{M}_{(2)}\mathbf{M}_{(3)}$ are:

\begin{equation}
\left\{
\begin{array}{l}
m_{11}=\cos(k_{0}\tilde{n}l)-k_{0}\tilde{n}\mathcal{B}\sin(k_{0}\tilde{n}l)\\[2mm]
m_{12}=-\frac{\mathrm{i}}{\tilde{n}}\sin(k_{0}\tilde{n}l)-\mathrm{i}k_{0}\mathcal{B}\cos(k_{0}\tilde{n}l)\\[2mm]
m_{21}=-\mathrm{i}k_0\mathcal{A}\cos(k_{0}\tilde{n}l)-\mathrm{i}\tilde{n}\sin(k_{0}\tilde{n}l)\\[2mm]
m_{22}=\cos(k_{0}\tilde{n}l)-\frac{k_0}{\tilde{n}}\sin(k_{0}\tilde{n}l)\mathcal{A}\\[2mm]
\end{array}
\right.
\end{equation}
With Equ.\ref{equ:sample} and Equ.\ref{equ:ref}, we denote the vacuum as medium 1 so that $\tilde{p}_{1}=\tilde{p}_{l=4}=1$.
The transmission coefficient of the sample $\tilde{t}$ is
\begin{equation}
\begin{aligned}
\tilde{t} &= {\tilde{t}_{{\rm{Sample}}}}/{\tilde{t}_{{\mathop{\rm Re}\nolimits} {\rm{f}}}}\\
 &= \frac{{2\exp \left( { - {\rm{i}}{k_0}d} \right)}}{{\left( {2 - {\rm{i}}{k_0}\mathcal{B}} \right)\cos \left( {{k_0}\tilde{n}l} \right) - \left( {{k_0}\mathcal{B}\tilde{n} + \frac{{\rm{i}}}{{\tilde{n}}} + {\rm{i}}\tilde{n}} \right)\sin \left( {{k_0}\tilde{n}l} \right) - {k_0}\left[ {\frac{1}{{\tilde{n}}}\sin \left( {{k_0}\tilde{n}l} \right){\rm{ + i}}\cos \left( {{k_0}\tilde{n}l} \right)} \right]\mathcal{A}}}
\end{aligned}
\end{equation}
Hence, $\Delta\tilde{n}$ can be solved as an unkown of a quadratic equation:
\begin{equation}
\begin{array}{l}

\frac{1}{2} (1- {\rm{e}}^{ - 2}){l_p}\Delta {\tilde{n}^2} + 2\left( {1 - {\rm{e}^{ - 1}}} \right){\tilde{n}}{l_p}\Delta \tilde{n}\\[2mm]
 + \tilde{n}^2{l_p} - \dfrac{{\tilde{t}\left[ {\left( {2 - {\rm{i}}{k_0}{l_p}} \right)\cos \left( {{k_0}{\tilde{n}}l} \right) - \left( {{k_0}{l_p}\tilde{n} + \frac{{\rm{i}}}{{\tilde{n}}} + {\rm{i}}\tilde{n}} \right)\sin \left( {{k_0}\tilde{n}l} \right)} \right] - 2\exp \left( { - {\rm{i}}{k_0}d} \right)}}{{\tilde{t}{k_0}\left[ {\frac{1}{{\tilde{n}}}\sin \left( {{k_0}\tilde{n}l} \right){\rm{ + i}}\cos \left( {{k_0}\tilde{n}l} \right)} \right]}} = 0\\[2mm]
\end{array}
\end{equation}
There exist two roots for the quadratic equation according to the quaratic formula. The way to pick a reasonable solution is to maintain the real part $\rm{Re}(\tilde{n}')=\rm{Re}(\tilde{n}+\Delta\tilde{n})\geqslant1$ and the imagine part of $\rm{Im}(\tilde{n}')=\rm{Im}(\tilde{n}+\Delta\tilde{n})>0$, for the calculated results should keep in line with the definition of physical quantities. The real part conductivity are shown as filled green cycles in FigS.\ref{Fig:S2}. The results of these two model agree well, while the multilayer model amplifies more effectively.
\vspace{\baselineskip} %

\noindent %
{\small
\lbrack 1\rbrack~M.~Born and ~E. Wolf,  \emph{Principles of optics: electromagnetic theory of propagation, interference and diraction of light} (Elsevier 2013).\\
}

\end{appendix}
\end{document}